\documentclass[aps,prc,twocolumn,showpacs,groupedaddress]{revtex4-1}
\usepackage{graphicx}  
\usepackage{bm}        
\usepackage{amssymb}   
\usepackage{soul} 
\begin{document}

\title{Angular dihadron correlations as an interplay between elliptic and
triangular flows}




\author{G.~Eyyubova}
\altaffiliation[Also at ]{
Skobeltsyn Institute of Nuclear Physics,
Moscow State University, RU-119991 Moscow, Russia
\vspace*{1ex}}
\affiliation{
Faculty of Nuclear Sciences and Physical Engineering, Czech Technical
University in Prague, CR-11519 Prague, Czech Republic
\vspace*{1ex}}
\author{V.L.~Korotkikh}
\affiliation{
Skobeltsyn Institute of Nuclear Physics,
Moscow State University, RU-119991 Moscow, Russia
\vspace*{1ex}}
\author{I.P.~Lokhtin}
\affiliation{
Skobeltsyn Institute of Nuclear Physics,
Moscow State University, RU-119991 Moscow, Russia
\vspace*{1ex}}
\author{S.V.~Petrushanko}
\affiliation{
Skobeltsyn Institute of Nuclear Physics,
Moscow State University, RU-119991 Moscow, Russia
\vspace*{1ex}}
\author{A.M.~Snigirev}
\affiliation{
Skobeltsyn Institute of Nuclear Physics,
Moscow State University, RU-119991 Moscow, Russia
\vspace*{1ex}}
\author{L.~Bravina}
\affiliation{
Department of Physics, University of Oslo, PB 1048 Blindern,
N-0316 Oslo, Norway
\vspace*{1ex}}
\author{E.E.~Zabrodin}
\altaffiliation[Also at]{
Skobeltsyn Institute of Nuclear Physics,
Moscow State University, RU-119991 Moscow, Russia
\vspace*{1ex}}
\affiliation{
Department of Physics, University of Oslo, PB 1048 Blindern,
N-0316 Oslo, Norway
\vspace*{1ex}}


\begin{abstract}
The hybrid model \textsc{hydjet++}, which considers soft
and hard processes, is employed for the analysis of dihadron angular 
correlations measured in Pb+Pb collisions at center-of-mass energy 
$\sqrt{s_{\rm NN}} = 2.76$ TeV. The model allows the study of both 
individual and mutual influence of lower flow harmonics, $v_2$ and 
$v_3$, on higher harmonics and dihadron angular correlations.
It is shown that the typical structure called {\it a ridge} in dihadron 
angular correlations in a broad pseudorapidity range could appear just
as interplay of $v_2$ and $v_3$. Central, semi-central and 
semi-peripheral collisions were investigated. Comparison of model 
results with the experimental data on dihadron angular correlations is 
presented for different centralities and transverse momenta $p_{\rm T}$.   
\end{abstract}

\maketitle

\section{Introduction}
\label{Introduction}

The measurement of azimuthal anisotropy and angular correlations of 
particles is an important tool for exploring properties of matter 
produced in nucleus-nucleus collisions. For non-central collisions of 
nuclei the initial azimuthal anisotropic overlap region leads to 
anisotropies in final particle distribution over the azimuth 
$dN/d\varphi$, which is characterized by the coefficients $v_n$ in the 
Fourier decomposition
\begin{equation}
\frac{dN}{d\varphi} \propto 1+ 2\sum_{n=1}^{\infty} v_n(p_{\rm T},\eta)
\cos \left[ n(\varphi-\Psi_n) \right] \ ,
\end{equation}
where $\Psi_n$ is the azimuth of the participant event plane of the 
$n$th harmonic, and coefficients $v_n$ depend on the transverse momentum 
$p_{\rm T}$ and pseudorapidity $\eta$. The two-particle angular 
correlation function, $C(\Delta\eta, \Delta\varphi)$, in the relative 
pseudorapidity $\Delta\eta = \eta^{\rm tr}-\eta^{\rm a}$ and the azimuth 
$\Delta\varphi = \varphi^{\rm tr}-\varphi^{\rm a}$ is sensitive to the 
collective flow of particles as well as to any other particle 
correlations in azimuthal angle and pseudorapidity. In the flow dominated 
regime the pair distribution can be expanded in a Fourier series:
\begin{equation}
\frac{dN^{pairs}}{d\Delta\varphi} \propto 1+ 2\sum_{n=1}^{\infty} 
V_n(p_{\rm T}^{\rm tr}, p_{\rm T}^{\rm a})\cos (n\Delta\varphi),
\label{Eq_Vn}
\end{equation}
where superscript indices refer to the two particles in a pair, usually 
called "trigger" and "associated" ones. The study of dihadron angular 
correlations in heavy ion collisions has revealed the new phenomena in 
collision dynamics, the so-called ridge and double-hump structure 
\cite{PHENIX_ridge,STAR_ridge,ALICE_ridge}. In order to explain this
correlation structure many mechanisms have been proposed, such as conical 
emission from either Mach-cone shock waves \cite{mach_cone,mach_cone2}
or Cerenkov gluon radiation \cite{Dr06}, large-angle gluon radiation, 
jets deflected by radial flow and path-length dependent energy loss 
(see \cite{qm08_ridge} and references therein). 

In Ref.~\cite{v3_Alver} the authors suggested that triangular flow 
might play an important role in the understanding of the ridge nature. 
Triangular flow, as well as higher flow harmonics, should arise due to 
initial state fluctuations in a collision geometry. Then, experiments 
at the CERN Large Hadron Collider (LHC) provided us with a new set of 
amazing results. Particularly, the ridge structure in a two-dimensional 
correlation function was also observed in proton-lead
\cite{CMS:2012qk,Aad:2012gla,Abelev:2012ola,Chatrchyan:2013nka} and in 
high multiplicity proton-proton collisions \cite{proton_ridge}. The 
origin of the ridge-like structure in $pp$ interactions and its 
similarity to that in Pb+Pb collisions are still open questions. 
Triangular flow measured in $p$+Pb reactions appeared to be compatible 
with $v_3$ in lead-lead collisions provided the multiplicity of 
secondary hadrons was the same. 
Traditionally proton-nucleus collisions are considered as cold nuclear 
matter effects, hence, the question is, Can the azimuthal anisotropy in 
cold nuclear matter have the the same strength as in hot nuclear matter?     

In heavy ion collisions the long-range, i.e., $|\Delta\eta|>2$, angular 
dihadron correlations at low and intermediate transverse momenta in 
(mid)central collisions were shown to be described with the sum of the 
Fourier harmonics $v_2 \div v _6$, found from independent flow analysis 
\cite{ALICE_ridge,ALICE_v3,ATLAS_ridge}. This implies that the $V_n$ 
coefficients in Eq.~(\ref{Eq_Vn}) factorize into two single-particle 
flow coefficients
\begin{equation}
\frac{dN^{pairs}}{d\Delta\varphi} \propto 1+ 2\sum_{n=2}^{\infty} 
v_n(p_{\rm T}^{\rm tr})v_n(p_{\rm T}^{\rm a})\cos (n\Delta\varphi).
\label{Eq_Vn_fact}
\end{equation}
The factorization was found to break at higher $p_{\rm T}$ and also for 
the first coefficient $V_1$ for the entire $p_{\rm T}$ range 
\cite{ALICE_ridge,ATLAS_ridge}. 
Are all of the six harmonics equally important for the description
of long-range correlations?

In the present paper we are going to study the role of only elliptic 
$v_2$ and triangular $v_3$ flows in the formation of long-range 
correlations. For this purpose we employ the \textsc{hydjet++} model
\cite{HYDJET_manual}, which merges parametrized hydrodynamics with jets.
In addition to hard processes, the unique feature of the model is the 
possibility to switch on and off the elliptic and triangular harmonics 
in order to investigate both their individual contributions and the
result of mutual interplay to the considered phenomena. The dihadron 
correlation function $C(\Delta\eta, \Delta\varphi)$ in lead-lead 
collisions at $\sqrt{s_{\rm NN}}=2.76$ TeV is investigated. The 
appearance of higher order harmonics $V_{n}$,\ $n>3$ in the correlation 
function is checked, and the obtained results are compareded against 
the available experimental data.    

\section{HYDJET++ model}
\label{hydjet}

The basic features of \textsc{hydjet++} model are described in a manual 
\cite{HYDJET_manual}. The model combines two components corresponding to 
soft and hard processes. The parameter which regulates the contribution 
of each component to the total event is the minimal transverse momentum 
$p_{\rm T}^{\rm min}$ of hard scattering. The partons either produced at 
or quenched down to the momenta below $p_{\rm T}^{\rm min}$ are 
considered to be thermal ones. Such partons do not contribute to the
hard part. 

The hard part of the model is based on \textsc{pythia} \cite{PYTHIA} and 
\textsc{pyquen} \cite{PYQUEN} generators, which simulate parton-parton 
collisions, parton radiative energy loss, and hadronization. The soft 
part of the model has no evolution stage from the initial state until  
hadronization, but rather represents a thermal hadron production already 
at the freeze-out hypersurface in accordance with the prescriptions of
ideal hydrodynamics adapted from the event generator \textsc{fast mc} 
\cite{Amelin}.

Strength and direction of the elliptic flow $v_2$ are regulated in the
\textsc{hydjet++} by two parameters. Spatial anisotropy $\epsilon(b)$
represents the elliptic modulation of the final freeze-out hypersurface 
at a given impact parameter $b$, whereas momentum anisotropy $\delta(b)$
deals with the modulation of flow velocity profile.
Additional triangular modulation of the freeze-out hypersurface,
\begin{eqnarray}
\nonumber
R(\varphi, b) &\propto& \frac{\sqrt{1-\epsilon(b)}}{\sqrt{1+\epsilon(b)
\cos{2[(\varphi-\Psi_2])}}} \\
\nonumber
&\times& \{1+\epsilon_3(b) \cos{[3(\varphi-\Psi_3)]}\} \ ,
\end{eqnarray}
produces triangular flow $v_3$ \cite{HYDJET_flow, HYDJET_v6}. Here 
$\epsilon_3$ is the new anisotropy parameter. The reaction plane 
$\Psi_2$ is fixed to zero and the $\Psi_3$ plane is generated randomly 
on an event-by-event basis. Both planes do not depend on $p_{\rm T}$ 
and $\eta$. Thus, the two planes are uncorrelated in accordance with 
the experimental data. The recent version of \textsc{hydjet++} is tuned 
to describe data on lead-lead collisions at the LHC energies 
\cite{HYDJET_flow, HYDJET_v6}. 

\section{HYDJET++ and dihadron correlations}
\label{results}

The two-particle correlation function is defined as the ratio of pair 
distribution in the same event (signal) to the combinatorial pair 
distribution (background), where pairs are not correlated. In experiment 
the background function is usually constructed with pairs from mixed 
events. The ATLAS and ALICE collaborations use the following definition 
\cite{ALICE_ridge, ATLAS_ridge}: 
\begin{equation}
C(\Delta\eta, \Delta\varphi)\equiv \frac{d^2N^{\rm pair}}{d\Delta\eta 
d\Delta\varphi} = \frac {N^{\rm mixed}}{N^{\rm same}}\times 
\frac{d^2N^{\rm same}/d\Delta\eta d\Delta\varphi}{d^2N^{ \rm mixed}/
d\Delta\eta d\Delta\varphi} \ ,
\end{equation}
where $N^{\rm mixed}$ and $N^{\rm same}$ are the number of pairs in the 
mixed events and same event, respectively. A one-dimentional (1D) 
correlation function $C(\Delta\varphi)$ is obtained by integrating 
$C(\Delta\eta, \Delta\varphi)$ over the pseudorapidity range 
$\Delta\eta$. Another definition of the correlation function is used by 
the CMS Collaboration \cite{CMS_ridge}:
\begin{equation}
\frac{1}{N^{\rm tr}}\frac{d^2N^{\rm pair}}{d\Delta\eta d\Delta\varphi }
=B(0,0)\times \frac {S(\Delta\eta, \Delta\varphi)}{B (\Delta\eta, 
\Delta\varphi)} \ ,
\label{CMS_def}
\end{equation}
where $N^{tr}$ is the number of trigger particles, and the signal and 
background are:
$$
 \displaystyle S(\Delta\eta, \Delta\varphi)=\frac{1}{N^{\rm tr}}
\frac{d^2N^{\rm same}}{d\Delta\eta d\Delta\varphi }, \quad 
 \displaystyle B(\Delta\eta, \Delta\varphi)=\frac{1}{N^{\rm tr}}
\frac{d^2N^{\rm mixed}}{d\Delta\eta d\Delta\varphi} \ . 
$$
This definition depends on event multiplicity, since it involves the 
number of associated particles where the pair of particles comes with 
approximately the same $\eta$ and $\varphi$ angles, $B(0,0)$.  

The background can be constructed from two single-particle spectra, 
$d^2N^{\rm tr}/d\eta d\varphi$ and $d^2N^{\rm a}/d\eta d\varphi$. 
Instead of correlating every two particles in mixed events, one 
correlates the yields in the given two bins. The yield represents the 
average over many events; therefore, the EbE correlations are washed 
out and the yield of pairs for the background function would be:
$$
B(\Delta\eta, \Delta\varphi)= \int \frac{d^2N^{\rm tr}}{d\eta^{\rm tr} 
d\varphi^{\rm tr}} \frac{d^2N^{\rm a}}{d\eta^{\rm a} d\varphi^{\rm a}} 
\delta^{\rm tr}_{\rm a}d\eta^{\rm a}d\eta^{\rm tr} d\varphi^{\rm a}
d\varphi^{\rm tr} \ ,
$$
where $\delta^{\rm tr}_{\rm a}=\delta(\eta^{\rm tr}-\eta^{\rm a} -
\Delta\eta)\delta(\varphi^{\rm tr}-\varphi^{\rm a}-\Delta\varphi)$.
Due to the absence of any detector effects in the model, spectra 
$dN/d\varphi$ as well as a background function $B(\Delta\varphi)$ should 
be flat. Thus, for function $B(\Delta\varphi, \Delta\eta)$ we use only 
$dN/d\eta$ distribution and assume flat distribution over 
$\Delta\varphi$. 

Fourier harmonics $V_n$ from Eq.~(\ref{Eq_Vn}) are defined directly from 
the correlation function $C(\Delta\varphi)$:
\begin{equation}
V_n=\langle cos(\Delta\varphi) \rangle = \frac{\sum_i C_i
(\Delta\varphi_i)\cdot \cos(n\Delta\varphi_i)}
{\sum_i C_i(\Delta\varphi_i)} \ . 
\end{equation}
If the collective azimuthal anisotropy is the dominant mechanism of the
correlation at large $|\Delta\eta|$, then $V_n$ coefficients would 
depend on single-particle anisotropies $v_n$ similar to 
Eq.~(\ref{Eq_Vn_fact}):
\begin{equation}
V_n(p_{\rm T}^{\rm low}, p_{\rm T}^{\rm low})=v_n(p_{\rm T}^{\rm low}) 
\times v_n(p_{\rm T}^{\rm low}) +\delta_n \ .
\label{Eq_ref_flow} 
\end{equation}
At low $p_T$ region the non-flow contribution $\delta_n$ is negligible,
thus leading to factorization of $V_n$. In experiment one usually 
defines the single-particle flow $v_n\{2PC\}$ via the two-particle 
correlation (2PC) function using $v_n$ at low $p_{\rm T}$ as a 
reference,
\begin{equation}
v_n\{{\rm 2PC}\}(p_{\rm T})=\frac{V_n(p_{\rm T}, p_{\rm T}^{\rm low})}
{v_n(p_{\rm T}^{\rm low})} \ ,
\label{Eq_2pc_flow}
\end{equation}
which effectively corresponds to two-particle cumulant method.  

\begin{figure*}
\includegraphics[width=0.32\textwidth]{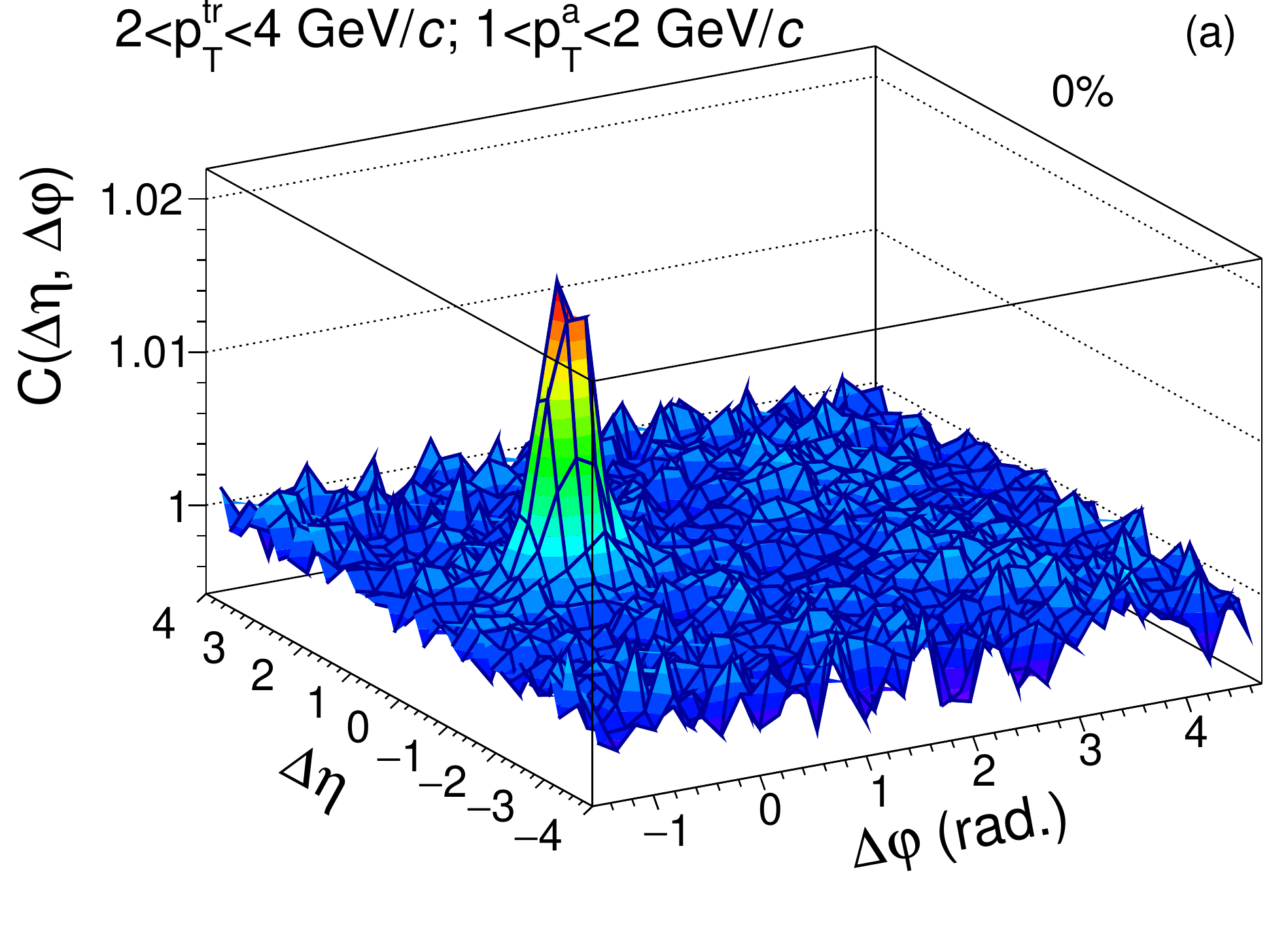} 
\hspace{1ex}
\includegraphics[width=0.32\textwidth]{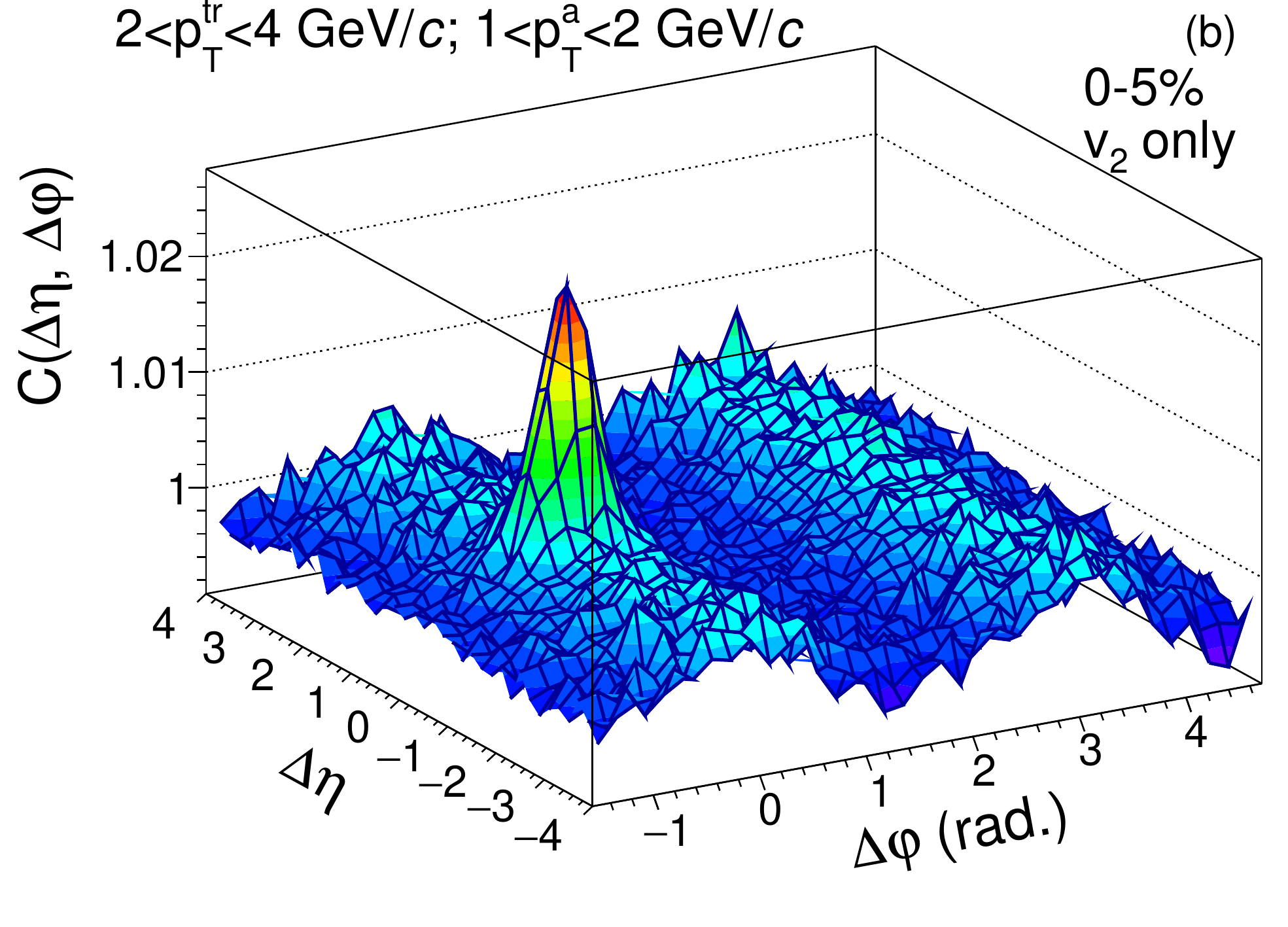}
\hspace{1ex}
\includegraphics[width=0.32\textwidth]{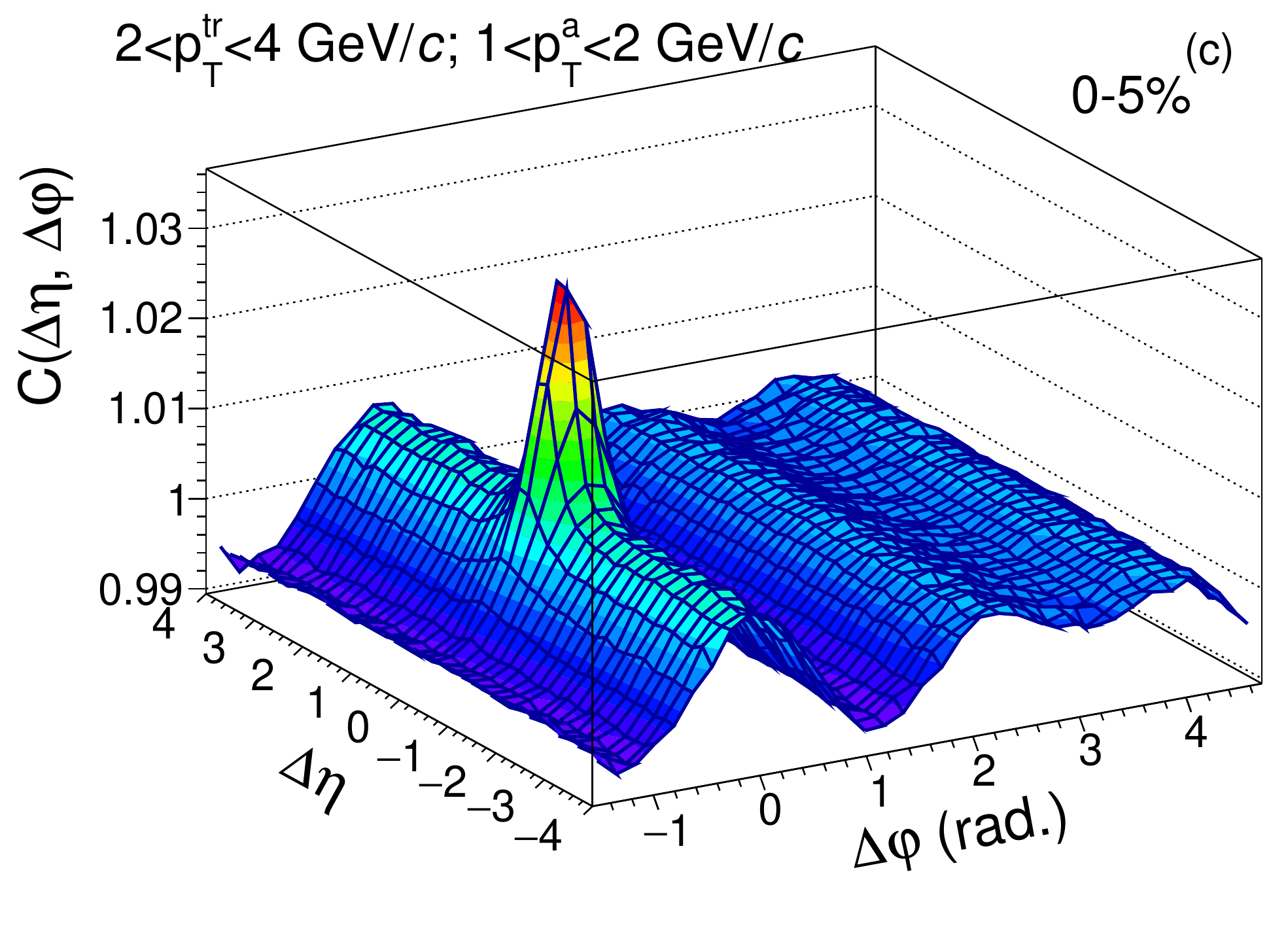}

\caption{(Color online)
Two-dimensional correlation function in \textsc{hydjet++} in Pb+Pb 
collisions at $\sqrt{s_{\rm NN}}=2.76$ TeV for $2<p_{\rm T}^{\rm tr}<4$ 
GeV/$c$ and $1<p_{\rm T}^{\rm a}<2$ GeV/$c$ for (a) central collisions 
with impact parameter $b=0 fm$, no flow, (b) centrality 0-5\% with only 
elliptic flow, and (c) centrality 0-5\% with both elliptic and 
triangular flow present.}
\label{ridge_hydjet}
\end{figure*}

Angular dihadron correlations contain all possible types of 
two-particle correlations. Many sources of two- or many-particle 
correlations, such as femtoscopic correlations, resonance decays, jets, 
and collective flow, are presented in the model. The long-range 
correlations over $\eta$ arise in the model merely due to collective 
flow. The correlation function $C(\Delta\eta, \Delta\varphi)$ calculated 
in \textsc{hydjet++} in the Pb+Pb collision at $\sqrt{s_{\rm NN}}=2.76$ 
TeV for $2<p_{\rm T}^{\rm tr}<4$ GeV/$c$ and $1<p_{\rm T}^{\rm a}<2$ 
GeV/$c$ is presented in Fig.~\ref{ridge_hydjet} for the cases of (a) 
absence of collective flow at zero impact parameter, (b) centrality 
0-5\%, only elliptic flow $v_2$ is turned on, and (c) centrality 0-5\%, 
both elliptic and triangular flow are present. The generated statistics 
are about $10^4$, $10^4$, and $10^5$ events, respectively. 
Figure~\ref{ridge_hydjet}(a) shows that the jet peak is highly 
suppressed at the away-side $\Delta\varphi \approx \pi$ due to jet 
quenching. Although remnants of it can be seen over a broad $\Delta\eta$ 
range at the away-side, no long-range azimuthal correlations are seen at 
the near-side. The long-range azimuthal correlations start to appear at 
the near-side in the presence of elliptic flow with the characteristic 
$\cos(2\Delta\varphi)$ pattern. They are flat in relative pseudorapidity 
up to $\Delta\eta\approx 4$, which corresponds to a flat pseudorapidity 
shape of the collective flow in the model. Finally, triangular flow 
enhances these near-side correlations, often referred to as a ridge. 
It also modifies the away-side of the distribution by producing a 
double-hump structure distinctly seen in Fig.~\ref{ridge_hydjet}(c).

In the \textsc{hydjet++} model, $v_2$ and $v_3$ anisotropies are 
introduced at the stage of thermal freeze-out, by means of the space 
modulation of the freeze-out volume and additional modulation of the 
flow velocity profile for the elliptic flow only. Thus, the model is 
insensitive to different origins of anisotropy and to the evolution 
dynamics from the initial state to the freeze-out stage. It is tuned, 
however, to describe the coefficients $v_2$ and $v_3$ both at low and 
at intermediate transverse momenta, where the hadrons from fragmenting 
jets start to dominate the particle spectrum. 
The interplay between $v_2$ and $v_3$ in the final state leads to the 
appearance of higher order flow harmonics, which reasonably describe 
data at mid-central collisions \cite{HYDJET_flow}. 

\begin{figure}
\includegraphics[width=0.5\textwidth]{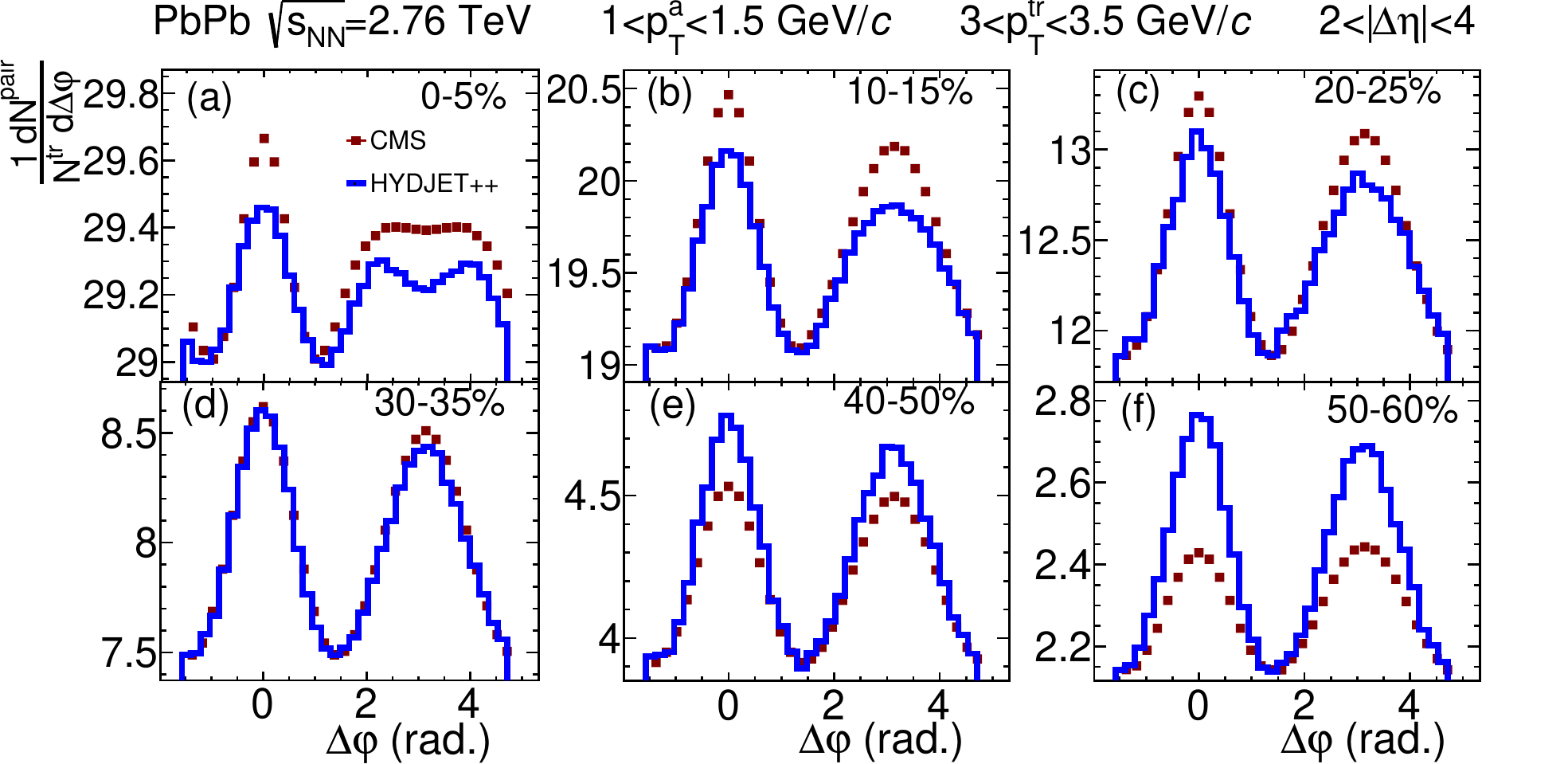}
\caption{(Color online) One-dimensional correlation function at 
$2<|\Delta\eta|<4$ in \textsc{hydjet++} in Pb+Pb collisions at 
$\sqrt{s_{\rm NN}}=2.76$ TeV for $3<p_{\rm T}^{\rm tr}<3.5$ GeV/$c$ 
and $1<p_{\rm T}^{\rm a}<1.5$ GeV/$c$ for different centralities in 
comparison with CMS data \cite{CMS_ridge}.
}
\label{C_phi_CMS}
\end{figure}

\begin{figure}
\includegraphics[width=0.5\textwidth]{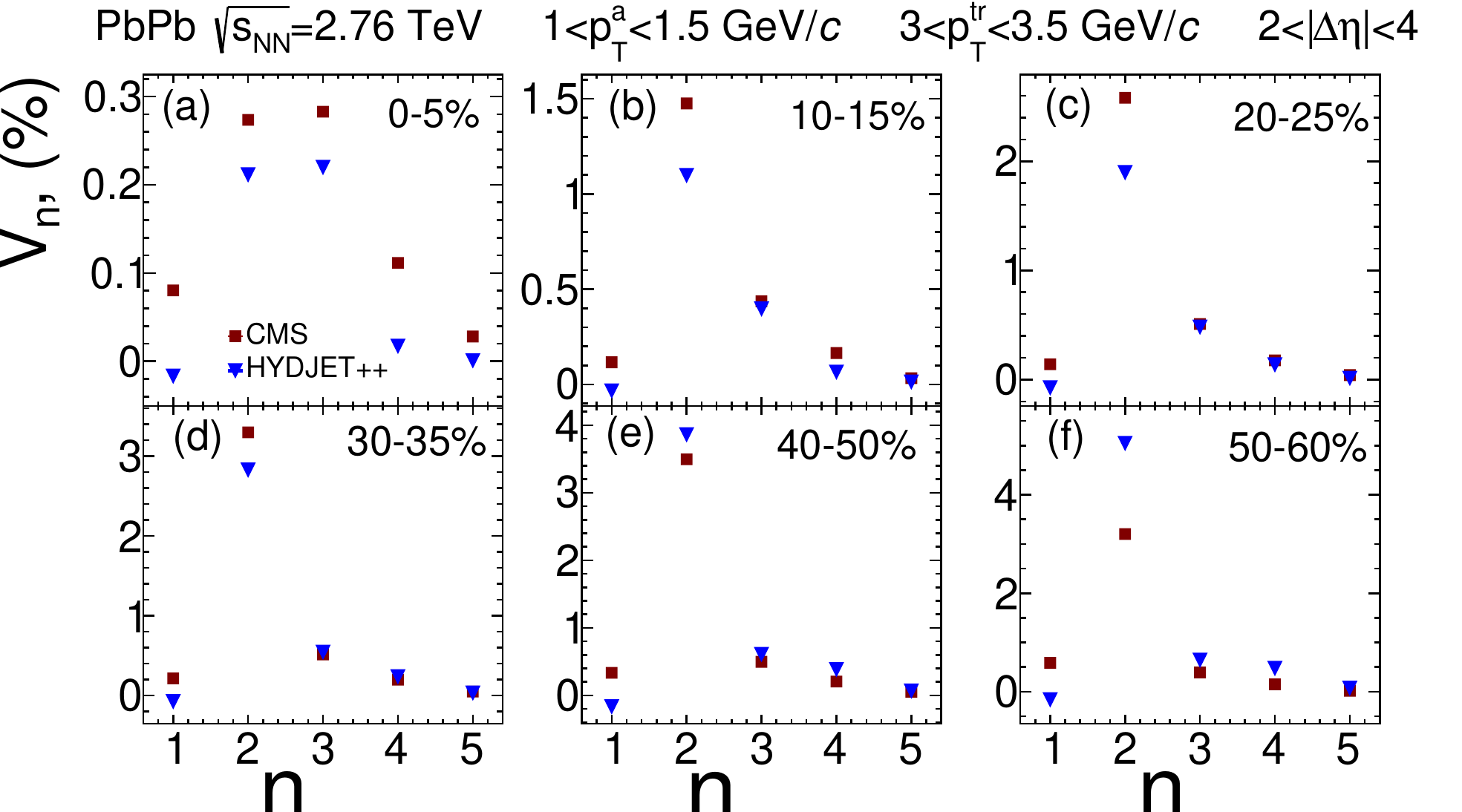}
\caption{(Color online) Fourier coefficients $V_n$ extracted for the 
1D correlation function, presented in Fig.~\protect\ref{C_phi_CMS}, in 
comparison with CMS data \cite{CMS_ridge}.}
\label{Vn_CMS}
\end{figure}

\begin{figure}
\includegraphics[width=0.5\textwidth]{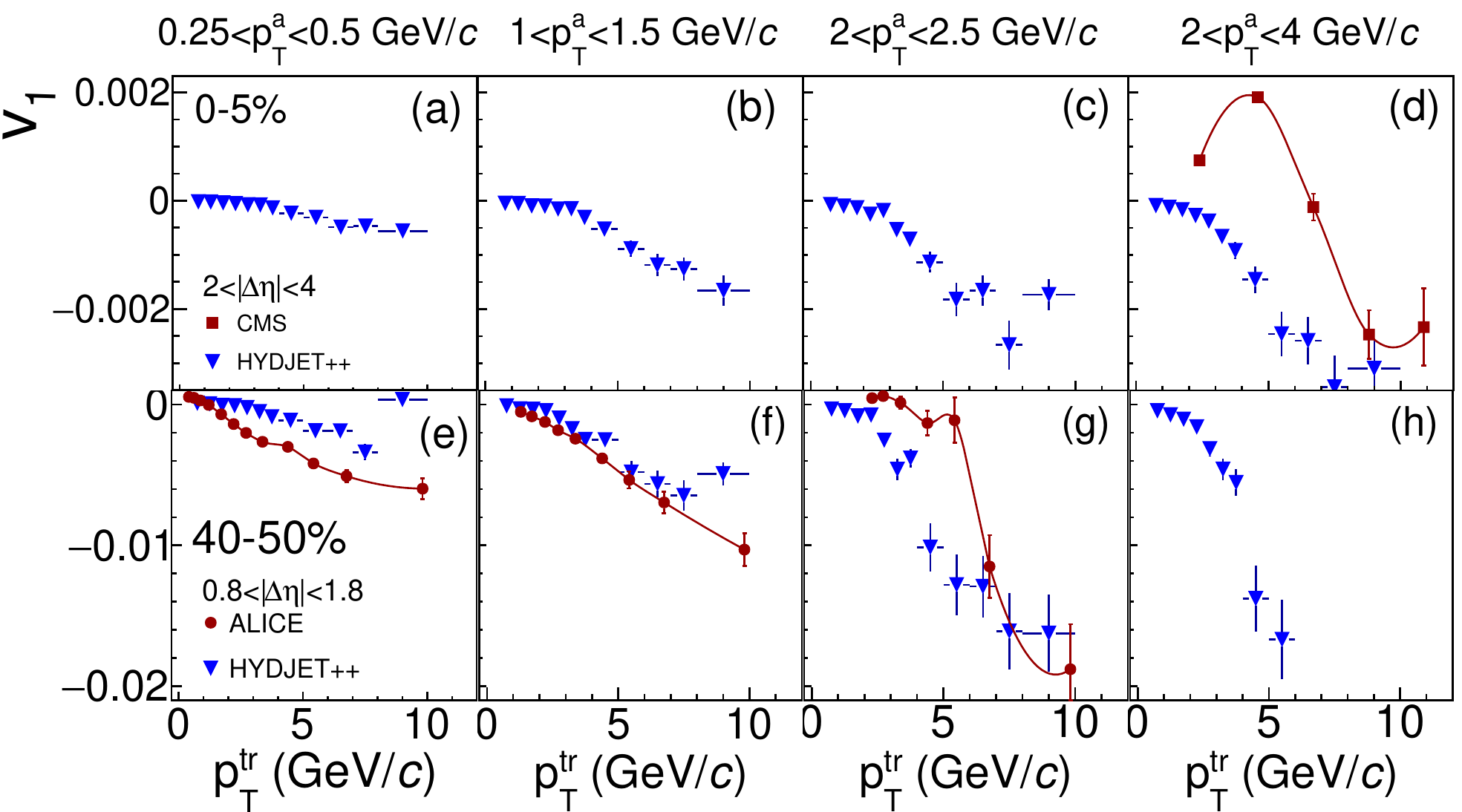}
\caption{(Color online) Upper row: Coefficients $V_1$ in the 0-5\% 
central Pb+Pb collision at $\sqrt{s_{\rm NN}}=2.76$ TeV for different 
associated $p_{\rm T}^{\rm a}$ as a function of $p_{\rm T}^{\rm tr}$ 
for \textsc{hydjet++} (circles) and CMS (squares) data 
\cite{CMS_dihadron}. 
Bottom row: The same as the top, but for centrality 40--50\%. Squares 
represent ALICE data from \cite{ALICE_v3}. Lines are drawn to guide
the eye.}
\label{fig_V1}
\end{figure}

\begin{figure}
\includegraphics[width=0.5\textwidth]{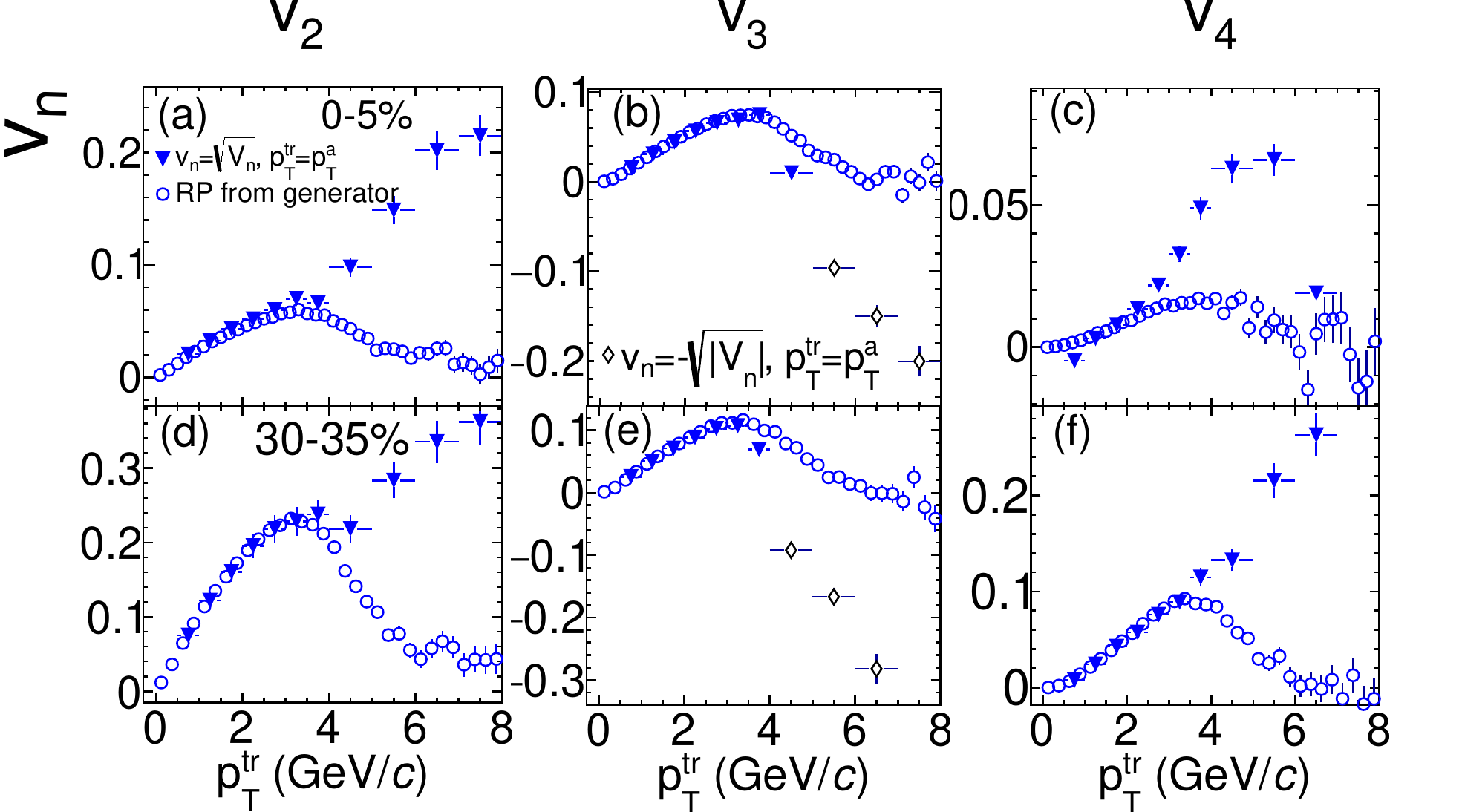}
\caption{(Color online) Upper row: Single-particle coefficients $v_n$ 
obtained with respect to a known reaction plane (open circles) and 
$v_n\{ {\rm 2PC} \}$ extracted from $V_n$ (full triangles down) for the 
same $p_{\rm T}^{\rm a}=p_{\rm T}^{\rm tr}$ in \textsc{hydjet++} generated 
0--5\% central Pb+Pb collisions at $\sqrt{s_{\rm NN}}=2.76$ TeV. Open
diamonds indicate points with negative $V_n$.
Bottom row: The same as the top, but for 30--35\% centrality.}
\label{fig_vn_fixpt}
\end{figure}

The results for long-range azimuthal correlations obtained with 
Eq.~(\ref{CMS_def}) for $1<p_{\rm T}^{\rm a}<1.5$ GeV/$c$ and 
$3<p_{\rm T}^{\rm tr}<3.5$ GeV/$c$ in \textsc{hydjet++} calculations
are plotted in Fig.~\ref{C_phi_CMS} onto the CMS data \cite{CMS_ridge} 
for different centralities. Since the correlation function given by
Eq.~(\ref{CMS_def}) depends on the multiplicity of associated particles, 
it does not always exactly coincide with the model. Therefore, 
\textsc{hydjet++} calculations are shifted on the constant value in such 
a way that the minima of $C(\Delta\varphi)$ in the data and in the model
coincide. In central collisions the model underestimates the data a bit
while in peripheral collisions the tendency is the opposite. The 
semi-central collisions are described quite well. Note that for 
centralities up to 35\% the difference between peak magnitudes of the 
\textsc{hydjet++} distributions and the experimental ones is less than 
3\%. It increases to 12\% in peripheral collisions with centrality 
50--60\%. To see the role of each of the Fourier coefficients $V_n$ more 
distinctly, we plot in Fig.~\ref{Vn_CMS} the values of the first five 
$V_n$ coefficients, calculated for the distributions shown in 
Fig.~\ref{C_phi_CMS}. At very 
central collisions all coefficients $V_n$ in the model are lower than 
those extracted from the data. At semi-central, semi-peripheral and even 
peripheral collisions all but $V_1$ and $V_2$ describe data rather well. 
At peripheral collisions $V_2$ in the model is higher than in data. This 
circumstance reflects the fact that the model predicts higher 
single-particle elliptic flow in the region of intermediate transverse 
momenta $3 < p_{\rm T} < 3.5$ GeV/$c$ compared to the data, see 
\cite{HYDJET_flow}, while the factorization holds.        
Note that there is no directed flow $v_1$ in the model, neither 
pseudorapidity odd $v_1$ nor even $v_1$, which is supposed to come from 
the initial state fluctuations as discussed in the literature 
\cite{v1_even}. Nevertheless, the $V_1$ component appears here due to 
violation of the momentum conservation, because in a part of the system 
with selected $p_T$ and $\eta$ ranges the momentum is not conserved. It 
was shown in \cite{Borgini_MomConserv} that the contribution of momentum 
conservation to the $V_1$ component can be presented by the term 
\begin{equation}
V_{1 m.c.}=-\frac{p_{\rm T}^{\rm tr} ~ p_{\rm T}^{\rm a}}
{M \langle p_{\rm T}^2 \rangle} \ ,
\end{equation}
where $M$ and $\langle p_{\rm T}^2\rangle$ are the multiplicity and 
average squared transverse momentum of the whole event, respectively. 
This approximation was made under assumption that the transverse 
momentum distribution is isotropic, or anisotropy is very weak and can 
be neglected. At higher $p_{\rm T}$ the cut on $\Delta\eta$ introduces 
additional unbalance in $\Delta\varphi$-distribution, since the 
near-side jet peak is almost completely eliminated by the cut, whereas 
the away-side jet peak stays partially. Figure~\ref{fig_V1} displays 
the $V_1(p_{\rm T}^{\rm tr})$ component calculated in \textsc{hydjet++} 
for different momenta of associated particles at two selected 
centralities. Available experimental data of ALICE and CMS 
collaborations are plotted onto the model results as well. It is 
clearly seen that the distributions can be approximated by a linear 
function only at a low-$p_T$ interval. Thus, the estimation of the 
$V_{1.m.c}$ contribution to the $V_1$ measured in the whole transverse 
momentum range requires additional study.   
Figure~\ref{fig_vn_fixpt} shows the coefficients $v_2\{2PC\}$, 
$v_3\{2PC\}$, and $v_4\{2PC\}$ extracted from $V_n$ at $|\Delta\eta|>2$ 
by Eq.~(\ref{Eq_ref_flow}) with $p_{\rm T}^{\rm tr}=p_{\rm T}^{\rm a}$ 
in comparison with $v_2$, $v_3$, $v_4$ calculated with respect to the 
known reaction plane at the generator level. In case of negative $V_n$ 
the coefficients $v_n\{2PC\}$ are taken as $v_n\{ {\rm 2PC} \} = 
-\sqrt{|V_n|}$. Comparison is presented for two centralities, 0--5\% 
and 30--35\%. One can see that in the range of $p_{\rm T}<3.5$ GeV/$c$ 
the 2PCmethod describes the $v_n$ coefficients pretty well. This means 
that $V_n$ coefficients factorize in this region into a product of two 
singular flow coefficients, and the collective flow is the dominant 
source of correlation. At higher transverse momenta, $p_{\rm T}>3.5$ 
GeV/$c$, the non-flow contribution of the jet component dominates. This 
contribution is negative for odd coefficients $v_n\{ {\rm 2PC} \}$.

\begin{figure}
\includegraphics[width=0.5\textwidth]{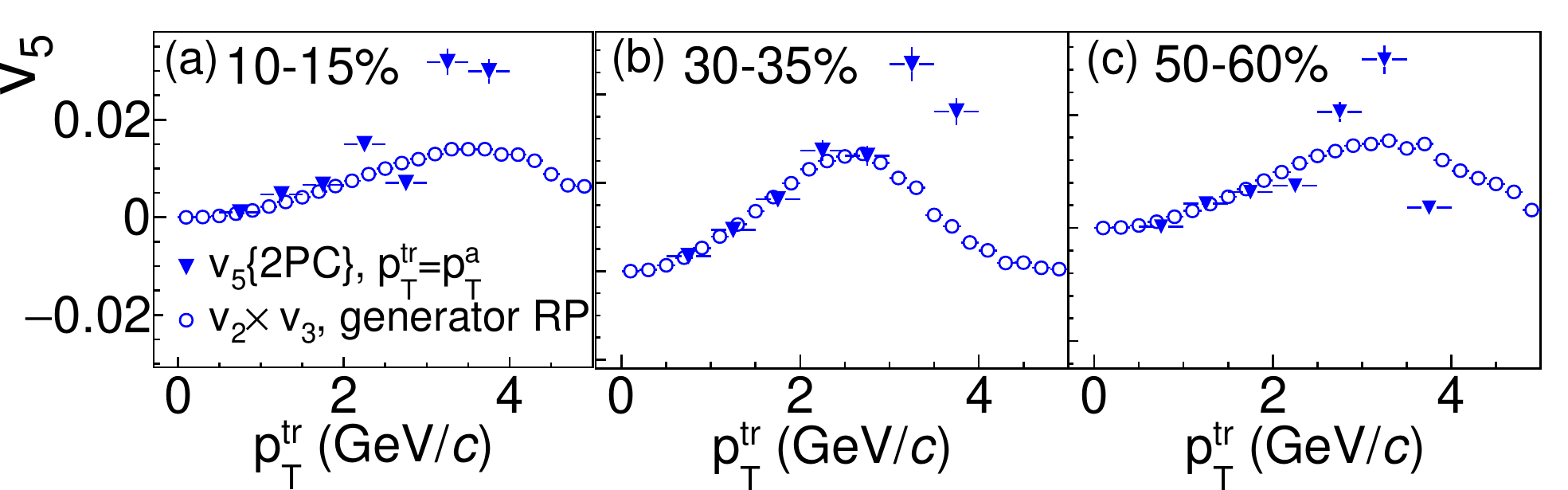}
\caption{(Color online) $v_5\{ {\rm 2PC} \}$ (triangles) obtained by 
Eq.~(\ref{Eq_ref_flow}) at centrality 10--15\% (left plot), 30--35\% 
(middle plot), and 50--60\% (right plot) in comparison to the product 
$v_2(p_T)\times v_3(p_{\rm T})$ (circles), obtained at generator levels 
with known reaction planes.}
\label{v5_fixpt}
\end{figure}

It is worth noting that higher order coefficients $V_n$,\ $n>4$ also 
appear in the model in $C(\Delta\varphi)$ decomposition, though they 
decrease rapidly with $n$, as shown in Fig.~\ref{Vn_CMS}. These 
coefficients at low $p_T$ can only originate from the lower order flow 
harmonics, $v_2$ and $v_3$. Figure~\ref{v5_fixpt} depicts pentagonal 
flow, $v_5\{2PC\}$, obtained by Eq.~(\ref{Eq_ref_flow}) at different 
centralities. The result is compared to the product $v_2(p_T)\times 
v_3(p_T)$, obtained at the generator level with a known reaction plane. 
It shows that the substantial contribution to $V_5$ comes from $v_2$ and 
$v_3$ harmonics at all centralities at $p_T \geq 1$ GeV/$c$. 

\section{Conclusions}
\label{concl}

The \textsc{hydjet++} model allows us to study the influence of a single 
harmonic, such as $v_2$ or $v_3$, as well as theirinterplay, on the 
final particle azimuthal distributions. This is the ideal situation, 
where all genuine higher-order initial fluctuations, which can distort 
the signal, are simply switched off. Elliptic flow contributes to all 
even harmonics of higher order, whereas the interplay of $v_2$ and $v_3$ 
leads to the appearance of odd harmonics in the model. In the present 
paper we obtained clear evidence that this mechanism allows one to 
describe also the dihadron correlations, including the double-hump 
structure, at mid-central collisions, where lower orders of collective 
flow dominates over higher harmonics. The measured amplitude of the 
ridge at mid-central collisions is well described by a superposition of 
elliptic and triangular flows. This is the main result of the paper.

Also, for pairs of particles with a large pseudorapidity gap 
($|\Delta\eta| > 2$) in a range of transverse momenta $p_{\rm T} < 3.5$ 
GeV/$c$, the coefficients $V_2$, $V_3$, and $V_4$ are found to factorize 
into the product of corresponding collective flow coefficients $v_n$ 
calculated in the model with a known reaction plane. In the absence of 
initial pentagonal fluctuations, pentagonal coefficient, $v_5\{2PC\}$, 
extracted from the dihadron correlation function follows approximately 
the scaling condition $v_5\{ { \rm 2PC} \} \propto v_2 v_3$ at $p_T 
\leq 2.5$ GeV/$c$ only.

\begin{acknowledgments}
We are grateful to L.~V.~Malinina for the fruitful discussions and 
valuable comments. This publication was supported by the Russian 
Scientific Fund under Grant No. 14-12-00110 in a part of computer 
simulation of 2D correlation functions in Pb+Pb collisions and 
extraction of the flow Fourier coefficients $V_{\rm n}$. G.E. 
acknowledges the European Social Fund within the framework of
realizing the project Support of Inter-sectoral Mobility and Quality
Enhancement of Research Teams at Czech Technical University in Prague,
CZ.1.07/2.3.00/30.0034.
\end{acknowledgments}

%
\end{document}